\documentclass[12pt]{article}
\usepackage{amssymb}

\input{tcilatex}
\newcommand{\da}{{\dot\a}}
\newcommand{\ad}{{\dot\a}}
\newcommand{\db}{{\dot\b}}

\newcommand{\btheta}{{\bar\theta}}

\makeatletter
\@addtoreset{equation}{section}
\makeatother
\renewcommand{\theequation}{\thesection.\arabic{equation}}

\def\ftoday{{\sl {Le \number\day \space\ifcase\month 
\or janvier\or f\'evrier\or mars\or avril\or mai
\or juin\or juillet\or ao\^ut\or septembre\or octobre
\or novembre \or d\'ecembre\fi\space \number\year}}}
\def\ptoday{{\sl {\number\day \space de\space \ifcase\month 
\or janeiro\or fevereiro\or mar{\c c}o\or abril\or maio
\or junho\or julho\or agosto\or setembro\or outubro
\or novembro \or dezembro\fi\space de\space \number\year}}}
\def\gtoday{{\sl {Den \number\day. \ifcase\month 
\or Januar\or Februar\or M\"arz\or April\or Mai
\or Juni\or Juli\or August\or September\or Oktober
\or November \or Dezember\fi\space \number\year}}}
\def\today{{\sl {\ifcase\month
\or January\or February\or March\or April\or May
\or June\or July\or August\or September\or October
\or November \or December\fi \space\number\day,\space 
                                            \number\year}}}
\newcommand{\journal}[4]{{\em #1~}#2\,(#3)\,#4}

\newcommand{\ijmp}{\journal {Int. J. Mod. Phys.}}

\newcommand{\cqg}{\journal {Class. Quantum Grav.}}

\newcommand{\np}{\journal {Nucl. Phys.}}
\newcommand{\pl}{\journal {Phys. Lett.}}
\newcommand{\mpl}{\journal {Mod. Phys. Lett.}}

\renewcommand{\a}{\alpha}
\renewcommand{\b}{\beta}
\newcommand{\g}{\gamma}

\renewcommand{\d}{\delta}         

\newcommand{\e}{\varepsilon}
\newcommand{\la}{\lambda}

\newcommand{\m}{\mu}

\newcommand{\p}{\psi}

\newcommand{\s}{\sigma}

\newcommand{\f}{{\phi}}

\newcommand{\LL}{{\cal L}}

\newcommand{\PP}{{\cal P}}

\newcommand{\es}{\\[3mm]}

\newcommand{\sla}{\raise.15ex\hbox{$/$}\kern -.57em}
\newcommand{\Sla}{\raise.15ex\hbox{$/$}\kern -.70em}

\newcommand{\lp}{\left(}
\newcommand{\rp}{\right)}

\newcommand{\identity}{{\bf 1\hspace{-0.4em}1}}
\newcommand{\complex}{{\kern .1em {\raise .47ex
\hbox {$\scriptscriptstyle |$}}
    \kern -.4em {\rm C}}}
\newcommand{\real}{{{\rm I} \kern -.19em {\rm R}}}
\newcommand{\rational}{{\kern .1em {\raise .47ex
\hbox{$\scripscriptstyle |$}}
    \kern -.35em {\rm Q}}}
\renewcommand{\natural}{{\vrule height 1.6ex width
.05em depth 0ex \kern -.35em {\rm N}}}

\newcommand{\half}{\ddfrac{1}{2}}
\newcommand{\pa}{\partial}

\newcommand{\dpad}[2]{{\displaystyle{\frac{\partial #1}{\partial #2}}}}

\newcommand{\ddfrac}[2]{{\displaystyle{\frac{#1}{#2}}}}

\newcommand{\twiddle}{\lower.9ex\rlap{$\kern -.1em\scriptstyle\sim$}}

\newcommand{\equ}[1]{(\ref{#1})}
\newcommand{\eq}{\begin{equation}}
\newcommand{\eqn}[1]{\label{#1}\end{equation}}
\newcommand{\ba}{\begin{array}}
\newcommand{\ea}{\end{array}}
\begin{document}

{\hfill\parbox{45mm}{{ 
hep-th/0108028\\
UFES-DF-OP2001/1
}} \vspace{3mm}

\begin{center}
{{\LARGE \textbf{Mass of the Fayet Hypermultiplet \\
[3mm] Induced by a Central Charge Constraint}}}
\vspace{12mm}

{\large Sortelano Araujo Diniz$^{\mathrm{(a,b)1 }}$
and Olivier 
Piguet$^{\mathrm{(b)}}$\footnote{{Supported by the Conselho Nacional de
Desenvolvimento Cient\'{\i}fico e Tecnol\'{o}gico CNPq -- Brazil.}}} 
\vspace{4mm}

$^{\mathrm{(a)}}$ \textit{Centro Brasileiro de Pesquisas F\'\i sicas (CBPF) 
\\[0pt]
Coordena\c c\~ao de Teoria de Campos e Part\'\i culas (CCP)\\[0pt]
Rua Dr. Xavier Sigaud 150 - 22290-180 - Rio de Janeiro - RJ - Brazil} 
\vspace{2mm}

$^{\mathrm{(b)}}$ \textit{Departamento de F\'\i sica, 
Universidade Federal do Esp\'\i rito Santo (UFES) 
\\[0pt]
Campus Universit\'ario de Goiabeiras - 29060-900 - Vit\'oria - ES - Brazil}

\vspace{4mm}

\texttt{E-mails: diniz@cce.ufes.br, piguet@cce.ufes.br}
\vspace{4mm}

{\it Revised version}
\bigskip
\end{center}

\begin{abstract}
We show that the mass of the Fayet hypermultiplet, which represents the
matter sector of N=2 supersymmetric Yang-Mills theory, may be induced
through a generalization of the central charge constraint usually proposed
in the literature. This mass showing up as a parameter of the supersymmetry
transformations, we conclude that it will stay unrenormalized at the quantum
level.
\end{abstract}


\section{Introduction}

In $N=2$ supersymmetric Yang-Mills theories, matter, i.e. spin 0 and $1\over2$
particles, is represented by the Fayet ``hypermultiplet''~\cite{fayet,West}.
It is represented in superspace by a certain constrained superfield. One
of these constraints, necessary in order to render finite the number of
local field components of the superfield, concerns the 
 central charge~\cite{Sohnius,West,Gaida,Hasler}. We  propose 
in the present paper a 
generalization of this constraint, involving a complex 
parameter $\la$ of mass 
dimension 1, which will appear in the supersymmetry transformation
rules of the component fields and show up in the resulting invariant
action as a contribution to the physical mass. 
  An interesting consequence of this construction is 
a nonrenormalization theorem for the mass, in models where the latter 
is entirely due to the constraint parameter $\la$, since the latter may
be defined as a parameter of the supersymmetry transformations laws.

 It is worthwhile recalling that 
the masses of the matter particles
in $N=2$ supersymmetric Yang-Mills theories may be generated by coupling the 
matter fields to a constant Abelian
super-Yang-Mills field strength, the values of the masses
being proportional to the value of this field 
strength~\cite{constant-f-str,constant-f-str'}. 
Thus, the mass generation via a generalized central charge constraint 
proposed in the present paper offers an alternative way of
generating the masses, with the parameter $\la$ replacing the constant
field strength. The interest of this alternative way is that it 
appears more natural, being purely algebraic.
In our scheme, indeed, $\la$ parametrizes the generalized central charge
constraint, which is algebraic.

In this  paper, we shall restrict ourselves to the construction of
the free theory of one hypermultiplet
in order to explain the mechanism in a simple
way. The construction of the full $N=2$ theory with coupling to a 
gauge supermultiplet is left for a forthcoming publication~\cite{diniz-g-p}.

The plan  is the following. After recalling, in Section 2, some
notations and definitions for $N=2$ super-Yang-Mills theory in the
superspace formalism, we give the construction of the Fayet matter
supermultiplet in Section 3, using the generalized supercharge 
constraint. We then construct the action in Section 4, showing the generation 
of the mass. The discussion of the nonrenormalization of this mass is
performed in Section 5.
Our conclusions are presented in Section 6.

\section{$N=2$ Superspace}

$N=2$ supersymmetry is defined by the Wess-Zumino superalgebra~\cite{West,
Wess} 
\begin{equation}
[\PP_{A},\PP_{B}]=T_{AB}^{C}\PP_{C}\ ,  \label{s-alg}
\end{equation}
where $\PP_{A}$ = 
$\{P_{a},\,Q_{\alpha }^{i},\,\bar{Q}_{i\dot{\alpha}
},\,Z,\,\bar Z\}$ is the set of infinitesimal generators: the
translations $P_{a}$, ($a=0,\cdots,3$), the supersymmetries 
$Q_{\alpha }^{i}$, $\bar{Q}{}_{i{\dot{\alpha}}}$ 
-- the  Lorentz spin indices $\alpha $ and ${\dot{\alpha}}$ taking the values
$1,2$, as well as the isospin SU(2) indices $i$ -- 
and the complex central charge $Z$, $\bar{Z}$.
Notations and conventions are explained in the Appendix.
Under the Lorentz transformations, $P$ transforms as a vector, $Q$ and 
$\bar{Q}$  as Weyl spinors -- in the Lorentz group representations 
$(\frac{1}{2},0)$ and $(0,\frac{1}{2})$, characterized by the 
indices $\alpha$ and $\dot{\alpha}$, respectively. $Z$ and $\bar{Z}$ 
are scalars. Moreover $Q$ and $\bar{Q}$ 
also transform as doublets of the isospin group SU(2) -- acting on the 
index $i$.

The generators $P$, $Z$ and $\bar Z$ are bosonic, whereas $Q$ and 
$\bar{Q}$ are fermionic. 
Accordingly, the bracket $[\cdot,\cdot]$ in the
l.h.s. of (\ref{s-alg}) is a graded commutator, i.e.
an anticommutator if both entries are fermionic,
and a commutator otherwise.

Finally, the structure constants of the superalgebra (\ref{s-alg}) -- the
``torsions'' -- are given by: 
\begin{equation}
T_\a^i{}_\b^j{}^z = 2i\e^{ij}\e_{\a\b}\ ,\quad 
T_\da^i{}_\db^j{}^{\bar z} = 2i\e^{ij}\e_{\da\db}\ ,\quad 
T_\a^i{}_\db^j{}^a = -2i\e^{ij}\s_{\a\db}^a\ ,
\label{free-torsions}
\end{equation}
all the other torsion coefficients vanishing.

\subsection*{Representation of the N=2 Wess-Zumino algebra:}

Our first task is to define the superspace representation of the Wess-Zumino
algebra (\ref{s-alg}). N=2 superspace \cite{West,Wess} is described by
the coordinates $\{\emph{X}^{A}\}=
\{x^{a},\theta _{\alpha }^{i},\bar{\theta}_{i\dot{\alpha}},z,\bar{z}\}.$ 
These coordinates are, respectively, the
space-time coordinates, Weyl isospin coordinates and its conjugate, complex
central charge and its conjugate\footnote{ See the Appendix
for our conventions of complex conjugation. 
As for the central charge coordinate, we have $\bar z$ $=$ $-z^*$, where 
$z^*$ is the complex conjugate of $z$.}. 
The spinor coordinates, $\theta $ and $\bar{\theta},$ 
are Grassmann (i.e. anticommuting or ''fermionic'') numbers,
the remaining ones are ordinary (i.e. commuting or ''bosonic'') numbers, so
the manifold coordinates satisfy the (anti)commutation rules:
\begin{equation}
[\emph{X}^A,\emph{X}^B] \equiv
\emph{X}^{A}\emph{X}^{B} - (-)^{ab}\emph{X}^{B}\emph{X}^{A} =0\ ,
\label{grading}
\end{equation}
where the grading $a=0$ if $\emph{X}^{A}$ is bosonic, and $a=1$ in the fermionic
case.

A superfield is a function in superspace, ${\phi }(\emph{X})$,
transforming under the generators of the superalgebra (\ref{s-alg}) as
follows: 
\begin{equation}
\begin{array}{l}
P_{a}{\phi }= \dfrac{\partial }{\partial x^{a}}\phi\ , \\[3mm] 
Q_\alpha ^i\phi =\left(\dfrac{\partial }{\partial\theta_i^a}
-i\pa_{\alpha \dot\beta}{\bar\theta}^{i\dot\beta}
+ \theta _{\alpha }^{i}\dpad{}{z}\right) {\phi}\ ,  \\[3mm] 
\bar{Q}{}_{i{\dot{\alpha}}}{\phi }=
\left( -{{\dfrac{\pa}{\pa{\bar{\theta}}^{i{\dot{\dot{\a}}}}}}}
+i\theta _{i}^{\a}\pa_{\a\da}
- {\bar{\theta}}_{i{\dot{%
\alpha}}}{{\dfrac{\partial }{\partial {\bar{z}}}}}\right) {\phi }\ ,  \\[3mm] 
Z{\phi }={{\dfrac{\partial }{\partial z}}}{\phi }\ ,\qquad
\bar{Z}{\phi }={{\dfrac{\partial }{\partial {\bar{z}}}}}{\phi }\ .
\end{array}
\label{transf-s-field}
\end{equation}
where we have defined 
\begin{equation}
\partial _{\alpha \dot{\alpha}}\equiv 
\sigma_{\alpha \dot{\alpha}}^{a}\partial _{a} \ ,\qquad
 {\bar\partial}_{\da\alpha}\equiv 
{\bar\s}_{\dot\alpha\alpha}^{a}\partial_{a} \ .
\label{d-slasch}\end{equation}
This provides the superfield representation of the superalgebra 
(\ref{s-alg}).

The covariant derivatives $D_{A}$ are superspace derivatives defined such
that $D_{A}{\phi }$ transform in the same way as the superfield ${\phi }$
itself. They are given by 
\begin{equation}
\begin{array}{l}
D_{a}{\phi } = \dfrac{\partial }{\partial x^{a}}\phi\ , \\[3mm] 
D_{\alpha }^{i}\phi =\left( {{\dfrac{\partial }{\partial \theta _{i}^{a}}}}%
+i \pa_{\alpha \dot{\beta}} {\bar{\theta}}^{i{\dot{\beta}}}
- \theta _{\alpha }^{i}\dpad{}{z}\right) {\phi }\ ,  \\[3mm] 
\bar{D}{}_{i{\dot{\alpha}}}{\phi }=\left( -{{\dfrac{\partial }{\partial {\bar{%
\theta}}^{i{\dot{\dot{\alpha}}}}}}}
- i\theta _{i}^{\alpha }\pa_{\a\da}
+ {\bar{\theta}}_{i{\dot{%
\alpha}}}{{\dfrac{\partial }{\partial {\bar{z}}}}}\right) {\phi }\ , \\[3mm] 
D_z{\phi }={{\dfrac{\partial }{\partial z}}}{\phi }\ , \qquad 
D_{\bar{z}}{\phi }={{\dfrac{\partial }{\partial {\bar{z}}}}}{\phi }\ .
\end{array}
\label{coderivative}
\end{equation}
and obey the same (anti)commutation rules as the generators, up to the signs
of the right-hand sides: 
\begin{equation}
\left[ D_{A},D_{B}\right] =-T_{AB}^{C}D_{C}\ ,  \label{alg-cov-der}
\end{equation}
the torsion coefficients $T$ being given in (\ref{free-torsions}).

The components of the supermultiplet corresponding to the superfield ${\phi }
$ are the coefficients of its expansion in powers of $\theta $ and ${\bar{%
\theta}}$. A generic component can be written as 
\begin{equation}
C_{n}=\left. (D)^{n}{\phi }\right| \ ,  
\label{gen-comp}\end{equation}
where $(D)^{n}$ is some product of $D_{\alpha }^{i}$ and ${\bar{D}}_{i{\dot{%
\alpha}}}$ derivatives, and where the symbol $|$ means that the expression
is evaluated at $\theta $ $=$ ${\bar{\theta}}=0$. It follows from this
remark and from the explicit transformation rules 
(\ref {transf-s-field}), that the action of the supersymmetry 
and central charge
generators on the components can be written as~\cite{Sohnius} 
\eq\ba{ll}
Q_{a}^{i}C_{n}= \left. D_{\alpha }^{i}(D)^{n}{\phi }\right| \ ,\quad 
&{\bar{Q}}_{i{\dot{\a}}}C_{n}= 
     \left. D_{i{\dot{\alpha}}}(D)^{n}{\phi}\right|  \ ,\es
ZC_n = \left. \pa_z (D)^{n}{\phi }\right| \ ,\quad
&\bar Z C_n = \left. \pa_{\bar z} (D)^{n}{\phi }\right| \ .
\ea\eqn{comp-transf}

\section{Construction of the Free Fayet Hypermultiplet}

The Fayet hypermultiplet 
\begin{equation}
\phi _{i}\equiv (\phi _{i},\chi _{\alpha },\bar{\psi}_{\dot{\alpha}},F_{i})
\label{Fayet}\end{equation}
is formed by two SU(2) doublets of complex scalars (${\phi }%
_{i},F_{i})$ and two Weyl spinors $(\bar{\psi}_{\dot{\alpha}},\chi _{a})$. 
It represents the matter sector of N=2 supersymmetric Yang-Mills
theories~\cite{West}, but we shall only consider the free Fayet hypermultiplet
in the present paper. It may be represented by an SU(2) doublet of complex
superfield\footnote{The same symbol ${\phi }$ represents 
the multiplet (\ref{Fayet}), the
corresponding superfield, as well as the first component of the latter, i.e.
its value at $\theta $ $=$ ${\bar{\theta}}$ $=0$.} ${\phi }_{i}(\emph{X})$
subjected to the supersymmetric constraints 
\begin{equation}
D_{\alpha }^{i}\phi ^{j}+D_{\alpha }^{j}\phi ^{i}=0\ ,\qquad 
 \bar{D}_{i\dot{\alpha}}\phi ^{j}+\bar{D}_{\dot{\alpha}}^{j}\phi _{i}=0\ .
\label{Fayet-constr}\end{equation}

\subsection*{Central charge constraint and supersymmetry transformations:}

The dependence of the superfield on the central charge coordinate $z$
leads in general to an infinity of local field components. 
In order to define a finite supersymmetry representation, one has to impose
a central charge constraint which restricts the dependence on $z$ and 
$\bar{z}$. We shall choose the constraint to be 
\begin{equation}
(\partial _{z} - e^{iw}\partial _{\bar{z}})\phi _{i}
=\lambda \phi _{i}\ ,\qquad
(\partial _{\bar{z}} - e^{-iw^{\ast }}\partial _{z})\bar{\phi}^{i}
= -\lambda ^{\ast }\bar{\phi}^{i}  \ .
\label{Znew-w}\end{equation}
It depends on a complex parameter $\lambda$ of the dimension
of a  mass
and on a dimensionless complex ``phase'' parameter $w$. 
This constraint
generalizes the one found in the literature~\cite{Gaida}, which 
corresponds
to zero $\lambda $ and $w$. We remark that, since 
$D_{\alpha}^{i}\ $\ and $\bar{D}_{\dot{\alpha}}^{i}$ 
commute with $\partial _{z}$ and 
$\partial _{\bar{z}}$ , the constraint above holds for the superfield 
$\phi_{i}(\bar{\phi}^{i})$ and all its derivatives, in particular on 
the derivatives which define the component fields 
\equ{gen-comp}.

In order to establish the supersymmetry transformation rules of the
hypermultiplet components, we first define the latters by the following
covariant derivatives of the superfield ${\phi }$: 
\begin{equation}
\begin{array}{l}
{\phi }_{i}\equiv \left. {\phi }_{i}\right| \ ,\quad \chi _{a}\equiv \left. 
\dfrac{1}{2\sqrt{2}}D_{\alpha }^{i}\phi _{i}\right| \ ,\quad {\bar{\psi}}_{{%
\dot{\alpha}}}\equiv \left. \dfrac{1}{2\sqrt{2}}{\bar{D}}_{{\dot{\alpha}}%
}^{i}{\phi }_{i}\right| \ ,\quad F^{j}\equiv \left. \dfrac{i}{8}D^{i\alpha
}D_{i\alpha }\phi ^{j}\right| =\left. \partial _{z}\phi ^{j}\ \right|\ .
\end{array}
\label{fayet-comp}
\end{equation}
The last equality in the definition of $F^{i}$ follows from the Fayet
constraints (\ref{Fayet-constr})  and the (anti)commutation rules
\equ{alg-cov-der}.

 Using formula (\ref {comp-transf}) together with the 
(anti)commutation rules \equ{alg-cov-der}, the Fayet 
constraints \equ{Fayet-constr} and the 
generalized constraint (\ref{Znew-w}), we can find
$(\lambda,w)$-dependent
transformation rules for the components, obeying the 
superalgebra \equ{s-alg}. However, before doing that, let
us observe that the {\it real part} of the "phase" $w$ may be eliminated 
through certain redefinitions.
Indeed, writing
\eq
w=u+iv\ ,\quad\mbox{with}\quad u,v\ \mbox{real}\ ,
\eqn{re-im-w}
we first note that the constraint (\ref{Znew-w}) reduces to the
form 
\begin{equation}
(\partial _{z} - e^{-v}\partial _{\bar{z}})\phi _{i}
=\lambda \phi _{i}\ ,\qquad
(\partial_{\bar{z}} - e^{-v}\partial _{z})\bar{\phi}^{i}
= -\lambda ^{\ast }\bar{\phi}^{i}  \ , 
\label{Znew}\end{equation}
if one redefines the central charge coordinates 
and the parameter $\lambda$ according to
\eq
z \to  e^{iu/2} z\ ,\quad {\bar z} \to e^{-iu/2} z\ ,\quad
\la \to e^{-iu/2}\ .
\eqn{redef-z}
However, compatibility of this redefinition with the covariant derivative
superalgebra of \equ{alg-cov-der} implies that the remaining
superspace coordinates has to be redefined, too:
\eq
\theta \to e^{iu/4} \theta\ ,\quad \btheta \to  e^{-iu/4} \btheta\ ,
\quad x \to x\ .
\eqn{redef-theta-x}
This amounts to a redefinition of the supercovariant derivatives as
\eq
{D}_\a^i \to e^{-iu/4}{D}_\a^i \ ,\quad
{{\bar D}}_{i\dot a} \to e^{iu/4}{\bar D}_{i\dot a}\ ,
\eqn{redef-D}
or, in view of \equ{fayet-comp}, to  phase redefinitions of the
component fields:
\eq\ba{l}
\f \to \f\ ,\quad  \chi \to e^{-iu/4}\chi \ ,\quad
\bar\psi \to e^{iu/4}\bar\psi \ ,\quad F \to e^{-iu/2} F\ , \es
\bar\f \to \bar\f\ ,\quad  \bar\chi \to e^{iu/4}\bar\chi \ ,\quad
\psi \to e^{-iu/4}\psi \ ,\quad \bar F \to e^{iu/2}\bar F\ .
\ea\eqn{field-redef}


Using now formula (\ref {comp-transf}) together with the Fayet 
constraints \equ{Fayet-constr} and the 
generalized central charge
constraint in the form (\ref{Znew}), we find the following 
$(\lambda,v)$-dependent
transformation rules for the components:
\begin{equation}
\begin{array}{ll}
Q_{\alpha }^{i}\phi _{j}=\sqrt{2}\delta _{j}^{i}\chi _{\alpha }\ ,\quad & 
\bar{Q}_{i\dot{\alpha}}\phi ^{j}=-\sqrt{2}\delta _{i}^{j}\bar{\psi}_{\dot{%
\alpha}}\ , \\[3mm] 
Q_{\a}^{i}\chi _{\b }=-\sqrt{2}i\varepsilon _{\a \b }F^{i}\ ,
& \bar{Q}_{i\dot{\alpha}}\chi _{\alpha }=-\sqrt{2}i\partial _{\alpha \dot{%
\alpha}}\phi _{i}\ , \\[3mm] 
Q_{\a }^{i}\bar{\psi}_{\dot{\beta}}=\sqrt{2}i\partial _{\a \dot{\beta}%
}\phi ^{i}\ , & \bar{Q}_{i\dot{\alpha}}\bar{\psi}_{\dot{\beta}}=e^{v}%
\sqrt{2}i\varepsilon _{\dot{\alpha}\dot{\beta}}(F_{i}-\lambda \phi _{i})\ ,
\\[3mm] 
Q_{\alpha }^{i}F_{j}=\sqrt{2}\delta _{j}^{i}\left( e^{-v}\partial _{\alpha 
\dot{\alpha}}\bar{\psi}^{\dot{\alpha}}+\lambda \chi _{\alpha }\right) \
,\quad & \bar{Q}_{i\dot{\alpha}}F^{j}=-\sqrt{2}\delta _{i}^{j}\partial
_{\alpha \dot{\alpha}}\chi ^{\alpha }\ ,
\end{array}
\label{Qtransfs}
\end{equation}
In the same way one finds the transformations for the conjugated
hypermultiplet $\bar{\phi}^{i} = (\bar{\phi}^{i},\bar{\chi}_{\dot{\alpha}%
},\psi _{\alpha },\bar{F}^{i})$: 
\begin{equation}
\begin{array}{ll}
Q_{\alpha }^{i}\bar{\phi}_{j}=\sqrt{2}\delta _{j}^{i}\psi _{\alpha }\quad & 
\bar{Q}_{i\dot{\alpha}}\bar{\phi}^{j}=\sqrt{2}\delta _{i}^{j}\bar{\chi}_{%
\dot{\alpha}} \\[3mm] 
Q_{\a }^{i}\bar{\chi}_{\dot{\beta}}=
- \sqrt{2}i\pa_{\a\dot{\b}}\bar{\phi}^{i}\quad 
& \bar{Q}_{i\dot{\alpha}}\bar{\chi}_{%
\dot{\beta}}=\sqrt{2}i\varepsilon _{\dot{\alpha}\dot{\beta}}\bar{F}_{i} \\%
[3mm] 
Q_{\a }^{i}\psi _{\b }=e^{v}\sqrt{2}i\varepsilon _{\a\b}
(\bar{F}^{i}-\lambda ^{\ast }\bar{\phi}^{i})\quad 
& \bar{Q}_{i\dot{\a}}\psi _{\alpha }
= -\sqrt{2}i\pa_{{\alpha}\dot\alpha}\bar{\phi}^{i} \\[3mm] 
Q_{\alpha }^{i}\bar{F}_{j}
= \sqrt{2}\d_{j}^{i}\partial_{\a\dot{\a}}\bar{\chi}^{\dot{\a}}\quad 
& \bar{Q}_{i\dot{\alpha}}\bar{F}^{j}
= \sqrt{2}\delta _{i}^{j}(e^{-v}
{\bar\pa}_{\dot{\alpha}\alpha }\psi ^{\alpha }
 + \lambda ^{\ast }\bar{\chi}_{\dot{\alpha}})
\end{array}
\label{free-tr-rules} 
\end{equation}
Finally, the central charge tranformation laws are
\eq\ba{ll}
Z \phi ^{i} = F^{i}\ ,
  &\bar{ Z} \phi^{i} = e^{v}(F^i - \lambda \phi^{i})\ , \es
Z \chi _{\alpha } = e^{-v}\partial _{\alpha \dot{\beta}}\bar{%
\psi}^{\dot{\beta}}+\lambda \chi _{\alpha }\ ,\qquad
  &\bar{ Z} \chi _{\alpha } =
       \partial_{\alpha \dot{\beta}}\bar{\psi}^{\dot{\beta}}  \ ,\es
Z \bar{\psi}_{\dot{\beta}} = 
    \partial_{\alpha \dot{\beta}}\chi^{\alpha }\ ,
  &\bar{ Z} \bar{\psi}_{\dot{\beta}}= e^{v}(\partial
   _{\alpha \dot{\beta}}\chi ^{\alpha }-\lambda \bar{\psi}_{\dot{\beta}})\ , 
\es
Z F^{i} = e^{-v}\square \phi ^{i} + \lambda F^{i} \ ,
  &\bar{ Z} F^{i}=\square \phi ^{i}\ ,
\ea\eqn{Ztransforma}
and similarly for the conjugate multiplet $\bar{\phi}^{i}$:
\eq\ba{ll}
\bar{ Z} {\bar \phi} ^{i} = -{\bar F}^{i}\ ,
  &Z  {\bar \phi^{i}} = -e^{v}({\bar F}^i-\lambda^* {\bar \phi}^{i})\ , \es
\bar{ Z}  {\bar \chi} _{\da} = -e^{-v}\partial_{\dot\alpha {\beta}}%
    \psi^{{\beta}} - \lambda^* {\bar \chi} _{\da}\ ,\qquad
  &Z  {\bar \chi}_{\da} =
      - \partial_{\da\b}{\psi}^{{\beta}}  \ ,\es
\bar{ Z} {\psi}_{\a} = 
    -\partial_{\a\db} {\bar \chi}^{\db}\ ,
  &Z  {\psi}_{\a} = -e^{v}(\partial_{\a\db}{\bar \chi}^{\db}
      - \lambda^* \psi_{\a})\ , \es
\bar{ Z} {\bar F}^{i} = -e^{-v}\square {\bar\phi}^{i}
     - \lambda^* {\bar F}^{i}\ ,
  &Z {\bar F}^{i} = - \square {\bar\phi}^{i}\ .
\ea\eqn{Zbarratransf.}  
  One readily checks that the equations 
(\ref{Ztransforma}) and (\ref{Zbarratransf.}) imply that for the
superfield $\phi$ ($\bar{\phi}$) or any of its components $C$
($\bar{C}$), the central charge constraints \equ{Znew} 
are satisfied, namely:
\begin{equation}
(\partial _{z}-e^{-v}\partial _{\bar{z}})C=\lambda C \ ,\qquad
 (\partial _{\bar{z}}-e^{-v}\partial _{z})\bar{C}
 =-\lambda ^{\ast } \bar{C}\ \   \label{DzDzX} \ ,
\end{equation}
 as it should be by construction.

Despite of the $(v,\lambda )$ - dependence of the 
transformations \equ{Qtransfs} - \equ{Zbarratransf.}, it is
simple to verify that the superalgebra algebra closes accordingly to 
\equ{s-alg}, independently of these parameters.
So, $v$ and $\lambda$ are completely free parameters, and 
 $\lambda$ may be complex.

\section{The Hypermultiplet Lagrangian}

In order to get the Lagrangian of the hypermultiplet constructed in the 
preceding section, we may use an
algorithm due to Hasler, based on the
\vspace{3mm}

\noindent{\bf Proposition}~\cite{Hasler}. 
Let be a superfield polynomial $L^{ij}$ 
-- called the ``kernel'' -- satisfying the conditions of zero symmetric
derivatives
\begin{equation}
D_{\alpha }^{(i}L^{jk)}=0\ , \qquad
\bar{D}_{\dot{\alpha}}^{(i}L^{jk)}=0\ .
\label{Hasler-cond}\end{equation}
Then the superfields
\begin{equation}
L\equiv -D_{k}^{\alpha }\Lambda _{\alpha }^{k}\ ,\qquad 
\bar{L}\equiv -D_{k\dot{\alpha}}\bar{\Lambda}^{k\dot{\alpha}}\   ,
\label{Hasler-L}\end{equation}
where
\begin{equation}
\Lambda _{\alpha }^{k}\equiv D_{i\alpha }L^{ik}\ ,\qquad 
\bar{\Lambda}^{k\dot{\alpha}}\equiv \bar{D}_{i}^{\dot{\alpha}}L^{ik} \ , 
\label{Hasler-lamb}\end{equation}
transform under supersymmetry -- with infinitesimal parameters 
$\xi$, $\bar\xi$ -- as
\begin{equation}\ba{l}
\delta L = i\pa_{z}
\lp \xi_i^\a\Lambda^i_\a+{\bar\xi}_{i\da}{\bar\Lambda}^{i\da}\rp
   -2i\partial_{a}
\lp{\bar\xi}_{i\da}{\bar\sigma}^{a\,\da\b}\Lambda^i_{\b}\rp\ ,\es
\delta \bar{L}=-i\pa_{\bar{z}}
\lp \xi_i^\a\Lambda^i_\a+{\bar\xi}_{i\da}{\bar\Lambda}^{i\da}\rp
   -2i\pa_{a}\lp \xi_i^\a \sigma^{a}_{\a\db}{\bar\Lambda}^{idb}\rp\ .
\ea\label{deltaL}\end{equation}
\vspace{3mm}

\noindent Let us apply this proposition to the 
kernel
\begin{equation}
L^{ij} = ie^{i\gamma}\pa_{\bar{z}}\bar{\phi}^{j}\phi ^{i}
+ i\bar{\phi}^{i} \pa_{z}\phi ^{j}\ ,  
\label{kernel-k}\end{equation}
where $\gamma$ is an arbitrary complex
''phase''\footnote{This phase slightly generalizes
the kernel proposed in the literature~\cite{West,Gaida}.}. The
conditions \equ{Hasler-cond} are satisfied due to the Fayet constraints
\equ{Fayet-constr} obeyed by both $\f^i$ and $\bar\f^i$.
We shall consider
a linear combination of the superfield defined by \equ{Hasler-L} and of
its complex conjugate: 
\begin{equation}
L_\theta \equiv  L  + e^{i\theta }\bar{L}  \ ,
\label{Lteta}\end{equation}
where $\theta$ is a complex number.
Since $\xi $ and $\bar{\xi}$ are independent parameters, it clearly
follows from \equ{deltaL} that
necessary and sufficient conditions for the supersymmetric variation
of $L_\theta$ to be a total divergence
are\footnote{Ref.~\cite{Hasler} considers the less general situation where
 $e^{i\theta}=\pm 1$ and 
$(\pa_{z}  -e^{i\theta}\pa_{\bar{z}})
( \xi_i^\a\Lambda^i_\a+{\bar\xi}_{i\da}{\bar\Lambda}^{i\da} )=0$.} 
\begin{equation}
(\partial _{z} - e^{i\theta }\partial _{\bar{z}})\Lambda _{\alpha }^{i}
=\mbox{total divergence}\ ,\qquad  
(\pa_{z} -e^{i\theta }\pa_{\bar{z}})\bar{\Lambda}_{\dot{\beta}}^{i}
= \mbox{total divergence}\ .  \label{totaldivergente}
\end{equation}
The kernel $L^{ij}$, and consequently $\Lambda $ and $\bar{\Lambda}$, 
being bilinear in $\phi^{i}$ and $\bar\phi^{i}$,
we easily check, using the equations (\ref{DzDzX}), (\ref{Hasler-lamb}), 
(\ref{Ztransforma}), (\ref{Zbarratransf.}) and (\ref{Znew}), 
that the condition (\ref{totaldivergente}) is satisfied if
\begin{equation}
 v=o\ ,\quad \theta = \gamma = 0\ ,\quad \lambda^* =-\lambda \ .
\label{restringeLamb}\end{equation}
We finally get a supersymmetric Lagrangian -- 
invariant up to a total derivative -- as: 
\eq
\LL=\dfrac{i}{24} L_\theta + \mbox{c.c.}
  = \LL_{\rm cin} + \LL_\la\ ,
\eqn{lagrangian1}
with 
\eq\ba{l}
\LL_{\rm cin} = \bar{F}F- \pa_{a}\bar{\phi}\pa^{a}\phi 
 - i\chi\sla\pa\bar{\chi} -i\psi\sla\pa\bar{\psi} \ , \es
\LL_\la = 
- \dfrac{\la}{2} \lp \bar F\f +i\bar\chi\bar\psi 
   + i\psi\chi - \bar \f F  \rp \ ,
\end{array}
\label{l-cin-l-lambda}
\end{equation}
where we have used the notation $(\sla\pa\bar\psi)_\a$ $=$ 
$\pa_{\a\db} {\bar\psi}^\db$ $=$
$\s^a_{\a\db}\pa_a{\bar\psi}^\db$.

 Let us note that the central charge constraint reads now 
\begin{equation}
(\partial _{z} - \partial _{\bar{z}})\phi _{i}
=\lambda \phi _{i}\ ,\qquad
(\partial_{\bar{z}} - \partial _{z})\bar{\phi}^{i}
= -\lambda ^{\ast }\bar{\phi}^{i}  \ , \quad
\mbox{with}\quad \lambda^*=-\lambda\ .
\label{Znewnew}\end{equation}
Although the transformation rules 
(\ref{Qtransfs}-\ref{Zbarratransf.}) constitute a representation of the
superalgebra \equ{s-alg} for any real $v$ and complex $lambda$, 
only the values $v=0$ and $lambda$ imaginary allow for an 
invariant Lagrangian.

We observe that the terms in $\la$ are mass terms, which have been induced
from the supersymmetry transformation rules we have defined
in the last Section. Of course,
it is still possible to add a mass term ``by hand''. This can be done
with Hasler's algoritm, too. 
We find in this way the invariant mass Lagrangian 
\eq  
\LL_\m = \m \lp \bar{F}\phi + \bar{\phi}F + i\bar{\chi}\bar{\psi} 
 - i\psi \chi \rp \ ,
\eqn{mass-lagr}
where $\m$ is a real mass parameter.

\subsection*{Equations of motions and masses:}

 The total Lagrangian reads
\eq
\LL_{\rm tot} = \LL_{\rm cin} + \LL_{\rm mass}\ ,
\eqn{tot-lagr}
where
\eq
\LL_{\rm mass} = M \lp \bar F\f +i\bar\chi\bar\psi \rp
   + M^* \lp  \bar \f F -i\psi\chi \rp\ ,\quad
\mbox{with } M = \m-\half\la   \ ,\  M^* = \m+\half\la\ ,
\eqn{gen-mass-lagr}
implying the equations of motions
\begin{equation}
\begin{array}{ll}
F_{i} + M \phi_{i}=0\ ,\qquad 
& \pa_{\a\da} \bar{\psi}^\da + M^* \chi_\a = 0\ ,\es
\square\phi_{i} + M^* F_{i} = 0 \ , \qquad 
& {\bar\pa}^{\da\a} \chi_\a - M \bar{\psi}^\da = 0\ .
\end{array}\label{motions}\end{equation}
These equations generalize the ones found in the
literature by the presence of the parameter $\lambda $, 
which contributes to the mass $M$. Despite of $\m$ and $\la$ being complex,
there is no tachyon in the theory. In fact, it is
simple to verify that all component fields satisfy Klein-Gordon 
equations with a real mass $|M|$:  
\begin{equation}
\left( -\square + \left|M\right|^{2}\right) {\varphi }=0\ ,\quad
 {\varphi } = \phi,\,F,\psi,\,\chi \ ,
\quad  |M|^2  =\mu^2 +\dfrac{1}{4}|\lambda|^2\ .
\end{equation}

\subsection*{ Parity:} 

 The mass Lagrangian \equ{gen-mass-lagr} in general is not invariant 
under parity invariance, defined by the 
transformations\footnote{The 4-component Dirac spinor 
$\Psi=(\chi_\a,\,\bar\psi{}^\ad)$ then
transforms as $\Psi$ $\to$ $\g^0\Psi$.
In terms of $\Psi$, the fermion mass term in\equ{gen-mass-lagr}
is a superposition of the scalar
$\bar\Psi\Psi$ and of the pseudoscalar $\bar\Psi\g^5\Psi$.} 
\eq\ba{l}
(x^0, {\bf x}) \rightarrow   (x^0, -{\bf x})\ ,\es
(\,\f^i\,,\         \chi^\a\,,\ \bar\psi{}_\db\,,\ F^i\,)  
\leftrightarrow
(\,\bar\f{}_i\,,\ \bar\chi_\da\,,\ \psi{}^\b\,,\ -\bar F_i\,)\ .
\ea\eqn{parity}
 This invariance however holds if (and only if) 
$M$ is purely imaginary, i.e.:
\eq
\m = 0 \ .
\eqn{parity-cond}

\section{Nonrenormalization of the Mass}

 The complex mass coefficient $M$ of
the general mass term \equ{gen-mass-lagr} has two contributions, 
namely one from the supersymmetry transformation parameter $\la$,
defined through the generalized central charge constraint (\ref{Znewnew}),
and the other one from
the free coefficient $\m$ of the separately invariant 
mass Lagrangian \equ{mass-lagr}. 
However, as we have seen, imposing invariance under parity 
implies that the mass is completely
determined by the parameter $\la$.
This means that, at the quantum level, where the symmetries are expressed by
Ward identities (see, e.g.,~\cite{alg-ren}), the total mass is then defined as a
parameter of the supersymmetry Ward identities -- to the contrary of the
usual case of a mass introduced as a separate 
invariant term of the action, such as
the term $\mathcal{L}_{\m}$ above. It follows that, in such a situation, the
mass \textit{is not renormalized}.

 However, for the latter result to hold, we must be certain that, after 
having set $\m=0$ in the action, a
counterterm of the form \equ{mass-lagr} will not appear as a radiative
correction. This can be guarantied by the presence of
 a ``protecting'' symmetry forbidding such a counterterm.
From the preceding discussion on parity, we can conclude that parity itself
provides a protecting symmetry. 
Parity invariance will thus assure
the absence of any independent mass counterterm. 

\vspace{3mm}

\noindent{\bf Remarks.}
\vspace{1mm}

\noindent 1.  We observe that the opposite situation, 
where the mass is entirely put by hand, namely the case 
$\la=0$, $\mu\not=0$, yields a Lagrangian equivalent to the one
discussed above and corresponding to $\la\not=0$, $\mu=0$. Indeed 
a field redefinition of the type \equ{field-redef} with
$u=\pi$ just corresponds to the interchange of the mass
Lagrangians $\LL_\la$ \equ{l-cin-l-lambda} and $\LL_\mu$ \equ{mass-lagr}.
Thus, even if the mass is put by hand, it is possible to reformulate
the theory in such a way that the mass is induced by the supersymmetry 
transformation laws, and thus would stay unrenormalized in presence of interactions.

\noindent 2.
Of course, our ``nonrenormalization theorem'' is trivial in the context of
the present free theory. It will however become relevant in the case of a
coupling with gauge fields~\cite{diniz-g-p}.

\section{Conclusions }

 We have shown that the constraint on the central charges of the 
$N=2$ hypermultiplet, necessary in order to keep the number of its
components finite, may be generalized introducing a dimensionful
complex parameter $\la$ and a dimensionless real parameter $v$ (Equ.
\equ{Znew}). 
These parameters modify the supersymmetry
transformation rules in a nontrivial way, but preserving the superalgebra 
\equ{grading}.
However, we were able to construct an invariant action only with these parameters 
restricted to $v=0$ and $\la$ purely imaginary. The latter parameter 
eventually contributes to the mass. 
A nonrenormalization theorem for the mass then follows if there is no other,
independent, contribution to it. This may be asssured 
thanks to a protecting symmetry, which turns out to be parity.
 Finally, we have noted  that, up to field redefinitions, the theory with mass
put by hand and the theory with mass
generated by supersymmetry transformation rules are equivalent.

Comparison with the results of~\cite{constant-f-str,constant-f-str'} 
is straightforward
at the level of the transformation laws. Performing the auxiliary field
redefinition $F$ $\to$ $F'$ $=$ $F+\frac{1}{2}\f$
in \equ{Qtransfs}, yields 
in the case of imaginary $\la$ the transformation laws corresponding 
to the equation (9) of Ref.~\cite{constant-f-str}, 
whith the constant field strength --
denoted there by $W_0$ -- being proportional to $\la$.

\subsubsection*{Acknowledgments}
We are much grateful to Richard Grimm for his help 
and very interesting discussions, in particuliar during his visit 
in Brazil, made possible by a financial support
of the CAPES/COPLAG, and during a month's stay of one of the authors
(O.P.) as Professeur Invit\'e at the Centre de Physique
Th\'eorique (CPT) of the Universit\'e de Aix-Marseille, France. 
{}O.P would like to warmly thank the members of the CPT for
their very kind invitation and for their hospitality.

\section*{Appendix. Notations and Conventions}

\renewcommand{\theequation}{\Alph{section}.\arabic{equation}} %
\setcounter{section}{1} 

Space-time is Minkovskian, 4-vector components are 
labelled by latin letters $a,b,\cdots$ 
$=$ $0,1,2,3$, the metric is choosen as  
\eq
\eta_{ab} = \mbox{diag} (-1, 1, 1, 1) \ .
\eqn{metric}
Weyl spinors are complex 2-component spinors $\p_{\a}$, $\a=1,2$, in the
$({1\over2},0)$ representation of the Lorentz group, or 
$\p_{\da}$, $\da=1,2$, in the $(0,{1\over2})$ representation. The $N=2$
internal symmetry group is ``isospin'' SU(2), isospinors being denoted
by $X^i$, $i=1,2$.

Isospin indices $i$ are raised and
lowered by the antisymmetric tensors $\e^{ij}$ and $\e_{ij}$:
\eq\ba{l}
X^{i}=\e^{ij}X_{j}\ ,\quad X_{i}=\e_{ij}X^{j}\ , \es
\mbox{with: \ }
\e^{ij}=-\e^{ji}\ ,\quad \e^{12}=1\ ,\quad \e_{ij}\e^{jk}=\d_i^k\ ,\quad
\e^{ij}\e_{kl}=\d_{l}^{i}\d_{k}^{j}-\d_{k}^{i}\d_{l}^{j}\ .
\ea\eqn{isospin-ind}
The same holds for the Lorentz spin indices, with the tensors $\e^{\a\b}$ and  
$\e^{\da\db}$ obeying to the same rules \equ{isospin-ind}.

Multiplication of spinors and isospinors is done, if not otherwise stated,
according to the convention
\eq
\p\chi = \p^\a\chi_\a\ ,\quad 
\bar\p\bar\chi = {\bar\p}_\da{\bar\chi}^\da\ ,\quad
UV = U^i V_i\ . 
\eqn{mult-conv} 

Our conventions for the complex conjugation, denoted by $^*$, are as follows:
\eq
(X^i_\a)^* = {\bar X}_{i\da}\ ,\quad 
({\bar X}_{i\da})^* = X^i_\a\ .
\eqn{compl-conj}

The matrices $\s^a$ and ${\bar\s}^a$ are defined by  
\eq\ba{l}
{\bar\s}^{a\,\da\a} = \e^{\a\b} \e^{\da\db} \s^a_{\b\db}\ ,\es
\s^0 = \identity \ ,\quad 
 \s^i\,(i=1,2,3)\, = \mbox{Pauli matrices} \ , \es
\bar\s{}^0 = \identity \ ,\quad 
 \bar\s{}^i\,(i=1,2,3)\, = -\mbox{Pauli matrices} \ , 
\ea\eqn{Pauli-matrices}
and obey the properties  
\eq
\s^a{\bar\s}^b + \s^b{\bar\s}^a = - 2\eta^{ab}\ ,\quad
\s^a_{\a\da} {\bar\s}_a^{\db\b} = -2\d^\b_\a \d^\db_\da\ .
\eqn{prop-P-matr}


\end{document}